\documentstyle[prl,twocolumn,aps,floats,epsf,psfrag,amsfonts]{revtex}

\tighten

\begin{document}
\draft

\twocolumn[\hsize\textwidth\columnwidth\hsize\csname @twocolumnfalse\endcsname
\title{Error correction for continuous quantum variables}

\author{Samuel L.~Braunstein}
\address{SEECS, University of Wales, Bangor LL57 1UT, UK}
\date{\today}
\maketitle

\begin{abstract}
We propose an error correction coding algorithm for continuous
quantum variables. We use this algorithm to construct a highly efficient
5-wavepacket code which can correct arbitrary single wavepacket errors.
We show that this class of continuous variable codes is robust against
imprecision in the error syndromes. A potential implemetation of the 
scheme is presented.
\end{abstract}
\vspace{3ex}
]

Quantum computers hold the promise for efficiently factoring large integers
\cite{Shor1}. However, to do this beyond a most modest scale they will 
require quantum error correction \cite{Shor2}. The theory of quantum 
error correction is already well studied in two-level or spin-$\case{1}{2}$
systems (in terms of qubits or quantum bits) 
\cite{Shor2,Steane,CS,Benn,Knill2,CRS}. Some of these results have been 
generalized to higher-spin systems \cite{Knill,Chau1,Chau2,Rains}. This
work applies to discrete systems like the hyperfine levels in ions but is 
not suitable for systems with continuous spectra, such as unbound wavepackets. 
Simultaneously with this paper, Lloyd and Slotine present the first treatment 
of a quantum error correction code for continuous quantum variables 
\cite{Lloyd}, demonstrating a 9-wavepacket code in analogy with Shor's
9-qubit coding scheme \cite{Shor2}. Such codes hold exciting prospects for
the {\it complete\/} manipulation of quantum systems, including both discrete 
and continuous degrees-of-freedom, in the presence of inevitable 
noise \cite{us2}.

In this letter we consider a highly efficient and compact error correction 
coding algorithm for continuous quantum variables. As an example, we 
construct a 5-wavepacket code which can correct arbitrary single-wavepacket 
errors. We show that such continuous variable codes are robust against 
imprecision in the error syndromes and discuss potential implementation 
of the scheme. This paper is restricted to 1-dimensional wavepackets 
which might represent the wave function of a non-relativistic 1-dimensional 
particle or the state of a single polarization of a transverse mode of 
electromagnetic radiation.  We shall henceforth refer to such descriptions 
by the generic term wavepackets \cite{fn1}.

Rather than starting from scratch we shall use some of the theory
that has already been given for error correction on qubits. In particular,
Steane has noted that the Hadamard transform
\begin{equation}
\hat H = {1\over\sqrt{2}}\left(\begin{array}{cr}
1 & -1 \\ 1 & 1 \end{array} \right) \label{Hadamard} \;,
\end{equation}
maps phase-flips into bit-flips and can therefore be used to form a class
of quantum error correction codes that consist of a pair of classical
codes, one for each type of `flip' \cite{Steane}. This mapping between
phase and amplitude bases is achieved with a rotation about the $y$-axis by 
$\pi/2$ radians in the Bloch sphere representation of the state. In
analogy, the position and momentum bases of a continuous quantum state
may be transformed into each other by $\pi/2$ rotations in phase-space.
This transition is implemented by substituting the Hadamard rotation in 
the Bloch sphere by a Fourier transform between position and momentum in 
phase-space. This suggests that we could develop the analogous quantum 
error correction codes for continuous systems \cite{newfn}.

We shall find it convenient to use a units-free notation where
\begin{eqnarray}
{\mbox{position}} &=& x \times ({\mbox{scale length}}) \nonumber \\
{\mbox{momentum}} &=& p\, /\, ({\mbox{scale length}}) \;,
\end{eqnarray}
where $x$ is a scaled length, $p$ is a scaled momentum and we have taken 
$\hbar=1$. (We henceforth drop the modifier `scaled'.) The position basis 
eigenstates $|x\rangle$ are normalized according to
$\langle x'|x\rangle = \delta(x'-x)$ with the momentum basis given by
\begin{equation}
|x\rangle = {1\over \sqrt{\pi}}\int dp \, e^{-2ixp}|p\rangle \label{momentum}\;.
\end{equation}
To avoid confusion we shall work in the position basis throughout
and so define the Fourier transform as an active operation on a state by
\begin{equation}
\hat{\cal F} |x\rangle = {1\over \sqrt{\pi}}\int dy \, e^{2ixy}|y\rangle 
\label{Fourier} \;,
\end{equation}
where both $x$ and $y$ are variables in the position basis. Note that
Eqs.~(\ref{momentum}) and (\ref{Fourier}) correspond to a change of
representation and a physical change of the state respectively.

In addition to the Fourier transform we shall require an analog to
the bit-wise exclusive-OR (XOR) gate for continuous variables. The XOR
gate has many interpretations including controlled-NOT gate, addition
modulo 2 and parity associated with it. Of these interpretations
the natural generalization to continuous variables is addition without
a cyclic condition. That is, we take
\begin{equation}
\begin{picture}(10, 5)
  \put(-2,9){\circle*{2.5}}
  \put(-2,9){\thinlines\line(0,-1){18}}
  \makebox(-4,-9)[c]{\large$\oplus$}
\end{picture}\!\!\!\!
|x,y\rangle = |x,x+y\rangle \label{XOR} \;.
\end{equation}
By removing the cyclic structure of the XOR gate we have produced a 
gate which is no longer its own inverse. Thus, in addition to the
Fourier transform and this generalized XOR gate we include their
inverses to our list of useful gates. This generalized XOR operation
performs translations over the entire real line, which are related to 
the infinite additive group on $\Bbb{R}$. The characters $\chi$ of this 
group satisfy the multiplicative property $\chi(x+y) = \chi(x)\chi(y)$
for all $x, y\in \Bbb{R}$ and obey the sum rule
\begin{equation}
{1\over\pi} \int_{-\infty}^{\infty} dx \,\chi(x) = \delta (x) \;,
\end{equation}
where $\chi(x) = e^{2ix}$. Interestingly, this sum rule has the same form 
as that found by Chau in higher-spin codes \cite{Chau2}. Once we have 
recognized the parallel, it is sufficient to take the code of a 
spin-$\case{1}{2}$ system as a basis for our continuous-variable code.

Based on these parallel group properties, we are tempted to speculate a 
much more general and fundamental relation: We conjecture that $n$-qubit 
error correction codes can be paralleled with $n$-wavepacket codes by
replacing the discrete-variable operations (Hadamard transform and XOR gate)
by their continuous-variable analogs (Fourier transform, generalized-XOR
and their inverses). As a last remark before embarking on the necessary
substitutions (in a specific example), we point out that the substitution
conjecture is only valid for qubit codes whose circuits involve only these
($\hat H$ and XOR) elements. We shall therefore restrict our attention 
to this class of codes.

\begin{figure}[thb]
\begin{psfrags}
\psfrag{p1}[c]{\Large $~~|\psi\rangle$}
\psfrag{z}[c]{\Large $|0\rangle$}
\psfrag{u1}[cb]{\footnotesize $~~~\,\hat H$}
\psfrag{u2}[cb]{\footnotesize $~~~~\,{\hat H}^\dagger$}
\epsfxsize=3.4in
\epsfbox[-60 -30 230 165]{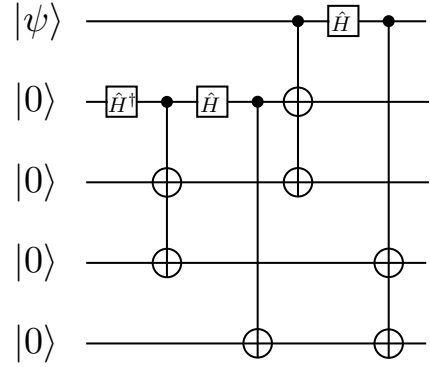}
\end{psfrags}
\caption{Quantum error correction circuit from \protect\onlinecite{SamSmol}.
The qubit $|\psi\rangle$ is rotated into a 5-particle subspace by the
unitary operations represented by the operations shown in this circuit.
Note that the 3-qubit gates are simply pairs of XORs.}
\label{fig1}
\end{figure}

An example of a suitable 5-qubit code was given by Laflamme et al.\
\cite{Laflamme}. We show an equivalent circuit in Fig.~\ref{fig1}
\cite{SamSmol}. As we perform the substitutions, we must determine which 
qubit-XOR gates to replace with the generalized-XOR and which with its 
inverse. To resolve this ambiguity, two conditions are imposed. First, we 
demand that the code retain its properties under the parity operation (on 
each wavepacket). We conclude that either gate may be chosen for the first 
operation on initially zero-position eigenstates. Ambiguity remains for the 
last four XOR substitutions. As a second step, the necessary and sufficient 
condition for quantum error correction \cite{Benn,Knill2}:
\begin{equation}
\langle x'_{\mbox{encode}}|\hat {\cal E}_\alpha^\dagger\, \hat {\cal E}_\beta
|x_{\mbox{encode}}\rangle = \delta(x'-x) \, \lambda_{\alpha\beta}\;,\;
\forall \;\alpha,\beta 
\label{cond} \;,
\end{equation}
must be met. Here $|x_{\mbox{encode}}\rangle$ encodes a single
wavepacket's position eigenstate in a multi-wavepacket state, 
$\hat {\cal E}_\alpha$ is a possible error that can be 
handled by the code and $\lambda_{\alpha\beta}$ is a complex constant 
independent of the encoded states. [Condition~(\ref{cond}) says that 
correctable errors do not mask the orthogonality of encoded states.]

\begin{figure}[thb]
\begin{psfrags}
\psfrag{p1}[c]{\Large $~~|\psi\rangle$}
\psfrag{z}[c]{\Large $|0\rangle$}
\psfrag{u1}[cb]{\small $~~~~\,\hat {\cal F}$}
\psfrag{u2}[cb]{\small $~~~~~\hat {\cal F}^\dagger$}
\psfrag{d}[bc]{\large $~~\dagger$}
\epsfxsize=3.4in
\epsfbox[-40 -10 210 165]{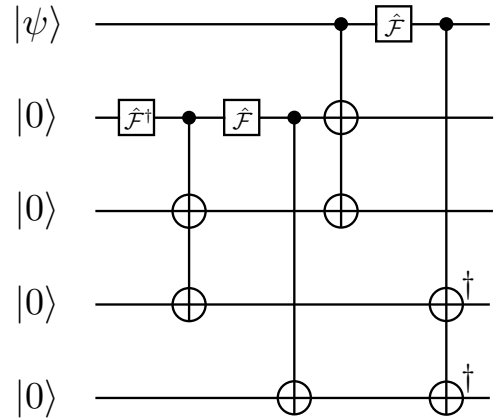}
\end{psfrags}
\caption{This `circuit' unitarily maps a 1-dimensional single-wavepacket 
state $|\protect\psi\protect\rangle$ into a 5-wavepacket error correction 
code. Here the auxiliary wavepackets $|0\protect\rangle$ are initially 
zero-position eigenstates. For degrees-of-freedom larger than qubits the 
ideal XOR is not its own inverse; here the daggers on the XOR gates 
represent the inverse operation.}
\label{fig2}
\end{figure}

In the case of a single wavepacket error, for our 5-wavepacket code,
it turns out that amongst the conditions of Eq.~(\ref{cond})
only $\langle x'_{\mbox{encode}}| \hat {\cal E}_{4\alpha}^\dagger\,
\hat {\cal E}_{5\beta}|x_{\mbox{encode}}\rangle$, having errors on
wavepackets 4 and 5, is affected by the ambiguity (see detail below). 
An explicit calculation of {\it all\/} the conditions shows that the 
circuit of Fig.~\ref{fig2} yields a satisfactory quantum error correction 
code (as do variations of this circuit due to the extra freedom with 
respect to the choice of operator acting on wavepackets 1-3). By 
analogy with the results for higher-spin codes, we know that this 
code is optimal (though not perfect) and that no four-wavepacket code 
would suffice \cite{Chau2}. The code thus constructed has the form
\begin{eqnarray}
|x_{\mbox{encode}}\rangle = {1\over\pi^{3/2}}
\int &&dw \,dy\,dz\, e^{2i(wy+xz)} \nonumber \\
&&\times|z, y+x, w+x, w-z, y-z\rangle \label{qecc} \,. \!\!\!\!
\end{eqnarray}

Let us demonstrate the calculation of one of the conditions specified
by Eq.~(\ref{cond}):
\begin{eqnarray}
&&\langle x'_{\mbox{encode}}| \hat {\cal E}_{4\alpha}^\dagger\,
\hat {\cal E}_{5\beta}|x_{\mbox{encode}}\rangle \\
&=&{1\over \pi^3}\int dw'\,dy'\,dz'\,dw\,dy\,dz\;
e^{2i(wy+xz-w'y'-x'z')} \nonumber\\
&&\times \delta(z'-z)\,\delta(y'-y+x'-x)\,\delta(w'-w+x'-x) \nonumber\\
&&\times\langle w'-z'|\hat {\cal E}_{\alpha}^\dagger|w-z\rangle
\langle y'-z'|\hat {\cal E}_{\beta}|y-z\rangle \nonumber \\
&=&{e^{-2i(x'-x)^2}\over \pi^3}\int dw\,dy\,dz\; e^{2i(x'-x)(w+y-z)}
\nonumber \\
&&\times \langle w-x'+x-z|\hat {\cal E}_{\alpha}^\dagger|w-z\rangle
\langle y-x'+x-z|\hat {\cal E}_{\beta}|y-z\rangle \,.  \nonumber 
\end{eqnarray}
Making the replacements $w\rightarrow w+z$ and $y\rightarrow y+z$ in
this last expression we obtain
\begin{eqnarray}
&=&{e^{-2i(x'-x)^2}\over \pi^3}\int dw\,dy\,dz\; e^{2i(x'-x)(w+y+z)}
\nonumber \\
&&\times \langle w-x'+x|\hat {\cal E}_{\alpha}^\dagger|w\rangle
\langle y-x'+x|\hat {\cal E}_{\beta}|y\rangle \\
&=&{\delta(x'-x)\over \pi^2}\int dw\,dy\,
\langle w|\hat {\cal E}_{\alpha}^\dagger|w\rangle\,
\langle y|\hat {\cal E}_{\beta}|y\rangle 
\equiv \delta(x'-x)\, \lambda_{\alpha\beta} \nonumber \;.
\end{eqnarray}
For the other cases we find by explicit calculation, for wavepackets
$j\ne k$, that
\begin{equation}
\langle x'_{\mbox{encode}}| \hat {\cal E}_{j\alpha}^\dagger\,
\hat {\cal E}_{k\beta}|x_{\mbox{encode}}\rangle = 
\delta(x'-x)\, \lambda_{\alpha\beta} \;.
\end{equation}
For $j=k$ this constant is found to be
\begin{equation}
\lambda_{\alpha\beta} ={C\over \pi^2}\int dw\, \langle w|
\hat {\cal E}_{\alpha}^\dagger\, \hat {\cal E}_{\beta}|w\rangle \;,
\end{equation}
where $C$ is formally infinite. 

\begin{figure}[thb]
\begin{psfrags}
\psfrag{state}[c]{$\hat{\cal E}(\hat p_1)|x,x,x\rangle$ \Huge $\{~~~~~~$}
\psfrag{z}[c]{\large $|0\rangle$}
\psfrag{d}[bc]{$~~\dagger$}
\psfrag{s1}[bc]{\large $s_1$} 
\psfrag{s2}[bc]{\large $s_2$}
\psfrag{s3}[bc]{\large $s_3$}
\epsfxsize=2.0in
\epsfbox[-160 -25 120 180]{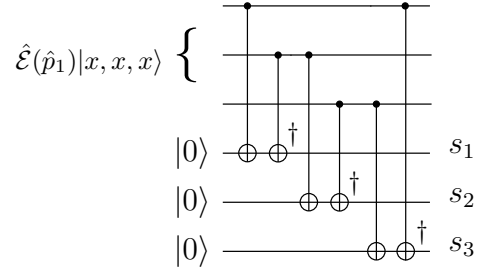}
\end{psfrags}
\caption{Syndrome calculation and measurement: A state with a single-wavepacket
position error (here on wavepacket $1$) enters and the differences
of each pair of positions is computed. The syndrome $\protect\{s_1,s_2,s_3
\protect\}$ may now be directly measured in the position basis.}
\label{fig3}
\end{figure}

We shall argue that this infinity
vanishes when the syndrome is read with only finite precision, which
is always going to be the real situation. However, this requires us to
demonstrate that our codes are robust: that for a sufficiently good
precision we may correct single-wavepacket errors to any specified 
accuracy. In order to understand how the error syndromes are measured,
let us consider a simpler code, namely, the continuous version
of Shor's original 9-qubit code:
\begin{eqnarray}
|x_{\mbox{encode}}\rangle=
{1\over \pi^{3/2}}\int &&dw\,dy\,dz\; e^{2ix(w+y+z)} \nonumber \\
&&\times|w,w,w,y,y,y,z,z,z\rangle \;,
\end{eqnarray}
where parity alone removes all ambiguity. (This code has 
been independently obtained by Lloyd and Slotine \cite{Lloyd}.)
Since this 9-wavepacket code corrects position errors and momentum errors 
separately, it is sufficient to study the subcode
\begin{equation}
|x_{\mbox{encode}}\rangle=|x,x,x\rangle \label{repcode} \;,
\end{equation}
designed to correct position errors on a single wavepacket. The most general 
position error (on a single wavepacket) is given by some function of the 
momentum of that system $\hat{\cal E}(\hat p)$ and need not be unitary 
on the code subspace [Eq.~(\ref{cond})]. The action of such an error 
on a wavepacket may be written in the position basis as
\begin{equation}
\hat{\cal E}(\hat p) |x\rangle = {1\over \pi}\!\int\! dy\,dp\; e^{2ip(y-x)}\,
{\cal E} (p) |y\rangle 
=\!\int\! dy\; \tilde {\cal E}(y) |x-y\rangle ,
\end{equation}
where $\tilde {\cal E}(x)$ is the Fourier transform of ${\cal E}(p)$. Thus 
the most general position error looks like a convolution of the wavepacket's
ket with some unknown (though not completely arbitrary) function. Suppose 
this error occurs on wavepacket one in the repetition code~(\ref{repcode}). 
Further, let us use auxiliary wavepackets (so-called ancillae) and compute 
the syndrome as shown in Fig.~\ref{fig3}, then the resulting state may 
be written:
\begin{equation}
\int dy\; \tilde {\cal E}(y)\, | x-y, x, x, -y, 0, y \rangle \;.
\end{equation}

Everything up till now has been unitary and assumed ideal. Now measure the
syndrome: Ideally it would be $\{-y, 0, y\}$ collapsing the wavepacket
for a specific $y$. Correcting the error is now easy, because we know 
the location, value and sign of the error. Shifting the first wavepacket
by the amount $y$ retrieves the correctly encoded state $|x,x,x\rangle$.
Note that this procedure uses only very simple wavepacket-gates: The
comparison stage is done {\it classically\/}, in contrast to the scheme 
of Lloyd and Slotine, where the comparison is performed at the amplitude 
level and involves significantly more complicated interactions \cite{Lloyd}.

It is now easy to see what imprecise measurements of the syndromes will
do. Suppose each measured value of a syndrome $s'_j$ is distributed randomly
about the true value $s_j$ according to the distribution
$p^{~}_{\mbox{meas}}(s'_j - s_j)$. We find two conditions for 
error-correction to proceed smoothly. First, $p^{~}_{\mbox{meas}}(x)$
must be narrow compared to any important length scales in
$\tilde{\cal E}(x)$. This guarantees that the chance for `correcting' the 
wrong wavepacket is negligible and reduces the position-error operator 
to an uninteresting prefactor. If the original unencoded state had been
$\int dx\, \psi(x)|x\rangle$ then after error correction we would obtain
the mixed state
\begin{eqnarray}
\int dx'\, dx\,dz \;&& \psi(x)\,\psi^\ast(x')\,
p^{~}_{\mbox{meas}}(z) \nonumber \\
&&\times |x-z,x,x\rangle\langle x'-z,x',x'| \label{corrstate} \;.
\end{eqnarray}
Thus, unless $p^{~}_{\mbox{meas}}(x)$ is {\it also\/} narrow compared to 
any important length scales in $\psi(x)$, decoherence will appear in the 
off-diagonal terms for wavepacket 1 of the corrected state~(\ref{corrstate}).
This second condition is also seen in the quantum teleportation of continuous 
variables due to inaccuracies caused by measurement \cite{us2}. These 
conditions roughly match those described by Lloyd and Slotine \cite{Lloyd}. 
We note that any syndrome imprecision will degrade the encoded states, 
though this precision may be improved by repeated measurements of the 
syndromes. For our 5-wavepacket example~(\ref{qecc}), syndromes consist of 
sums of two or more wavepacket positions or momenta and are measured 
similarly.

It should be noted that Chau's higher-spin code \cite{Chau2} could have 
been immediately taken over into a quantum error correction code for 
continuous quantum variables in accordance with our substitution
procedure. However, we have produced an equivalent code with a more 
efficient circuit prescription: Whereas Chau gives a procedure for 
constructing his higher-spin code using 9 generalized XOR operations,
the circuit in Fig.~\ref{fig2} requires only 7 such gates or their inverses. 
In fact, we could run this substitution backwards to obtain a cleaner 
5-particle higher-spin code based on Eq.~(\ref{qecc}).

In order to consider potential implementations of the above 
code let us restrict our attention to a situation where the
wavepackets are sitting in background harmonic-oscillator potentials.
By the virial theorem the form of a wavepacket in such a potential is 
preserved up to a trivial rotation in phase-space with time. The
two operations required may be implemented simply as follows: The
rotation in phase-space, Eq.~(\ref{Fourier}), may be obtained by 
delaying the phase of one wavepacket relative to the others, and the XOR
operation, Eq.~(\ref{XOR}), should be implemented via a quantum
non-demolition (QND) coupling. There exists extensive experimental 
literature on these operations both for optical fields and for trapped 
ions \cite{us2,QND1,QND2,QND3,QND4}.

The conjecture put forth in this letter leads to a simple, 2-step
design of error correction codes for continuous quantum variables.
According to this conjecture, any qubit code, whose circuit operations
include only a specific Hadamard transformation, its inverse and the ideal
XOR, may be translated to a continuous quantum-variable code, by substituting
these operators with their continuous analogs and then imposing two
criteria -- parity invariance and the error-correction condition -- which
remove any ambiguities in the choice of operators. We 
demonstrate the success of this coding procedure in two examples (one
based on Shor's 9-qubit code \cite{Shor2}, and a second based on a variation 
of the Laflamme et al.\ 5-qubit code \cite{Laflamme,SamSmol}).
The 5-wavepacket code presented here is the optimal continuous
encoding of a single 1-dimensional wavepacket that protects
against arbitrary single-wavepacket errors. We show that this code
(and in fact the entire class of codes derived in this manner) are
robust against imprecision in the error syndromes. The potential
implementation of the proposed class of circuits in optical-field and
ion-trap set-ups is an additional incentive for further investigation of
the robust manipulation of continuous quantum variables.

\vskip 0.2truein
This work was funded in part by EPSRC grant GR/L91344. The author
appreciated discussions with N.\ Cohen, H.\ J.\ Kimble, D.\ Gottesman 
and S.\ Schneider.


\begin{references}
\bibitem{fn1}Although we consider multi-wavepacket states as
one-dimensional we would not want them to physically overlap so they
could, for example, be displaced one from another in an orthogonal
direction.
\bibitem{newfn}An example using as few as 7 qubits to correct an 
arbitrary single-qubit error in a 1-qubit encoded state
can be found in~\onlinecite{Steane}.
\end{references}
\end{document}